\documentstyle[aps,preprint]{revtex}
\begin{document}
\draft

\title{
Long-range correlations in diffusive systems away from equilibrium
}

\author{I. Pagonabarraga and J.M. Rub\'{\i}}
\address{
Departament de F\'{\i}sica Fonamental, Facultat de F\'{\i}sica \\
Universitat de Barcelona\\
Diagonal 647, 08028 Barcelona, Spain
}
\date{\today}

\maketitle
\begin{abstract}

         We study the dynamics of density fluctuations in purely diffusive
systems away from equilibrium.  Under some conditions the static density
correlation function becomes long-ranged. We then analyze this behavior in the
framework of nonequilibrium fluctuating hydrodynamics.

\end{abstract}

\section{Introduction}

        Recently, there has been a great interest in the study of long-range
spatial correlations in non-equilibrium systems. A wide variety of models
governed by Langevin-like equations have been proposed. A first class of models
have been introduced to describe systems, in the presence of external noise
sources, in which the fields are conserved only in average.  As an example we
could mention models of sand-piles \cite{Grn} or diffusion in disordered media
\cite{Ara}-\cite{Rub}, which display long-range order. A second group of models
would account for situations in which internal noise sources are present and
the total values of the fields are conserved.  Among them one finds turbulence
models in fluids \cite{For}, continuum models for interface growing or directed
polymers \cite{Med} and some models for driven-diffusive systems
\cite{Grn},\cite{Gar}. In them, although fluctuations have an internal origin,
they are not necessarily related to dissipation. Whereas in the former
long-range behavior is always observed, in the latter long-range correlations
appear when some intrinsic anisotropy is present.

        Hydrodynamic systems constitute a large class of systems which present
also internal fluctuations coming from the presence of microscopic degrees of
freedom. The existence of a fluctuation-dissipation theorem introduces new
aspects in the behavior of the correlations. In particular, it is shown that
the form of the fluctuation-dissipation theorem is determinant in order to
produce such correlations. In this sense, a preliminary result was found for
thermal conducting systems \cite{Agu} where the temperature correlation
function decays as s$^{-1}$, s being the distance between the points in
consideration (see eq.(\ref{24})). Such long-range correlations have been
observed experimentally in liquids under thermal gradients \cite{Law}.

        Our purpose in this paper is to analyze the origin of long-range
correlations in purely diffusive systems, described only by one variable: the
number density of particles. In these systems diffusion is the only mechanism
responsible for the maintenance of non-equilibrium steady states. Therefore, we
are not considering the possibility of applying external forces ( as for
example
electric fields) which could be the origin of long-range correlations even if
the stationary state is homogeneous \cite{Gun}. The dynamics of fluctuations
around nonequilibrium steady states is dictated by nonequilibrium fluctuating
hydrodynamics. We can imagine a general situation in which the diffusion
coefficient may depend on the position. This is what happens for example when
diffusion takes place in a suspension of particles in the presence of
hydrodynamic interactions or in an inhomogeneous medium. In the first case the
diffusion coefficient is not constant due to the local concentration value,
taking into account the fact that the distance among the particles can hinder
their movement, while in the second the diffusion coefficient depends on the
position through the inherent properties of the medium where the particles are
diffusing. While in the first system we obtain long-range
correlations, in the second density fluctuations are delta correlated.

        The paper is organized as follows. In section 2 we introduce the
models. We specify the steady states and the fluctuation  evolution equations.
These results enable us, in section 3, to determine the static correlation
functions.  Finally, in section 4, we discuss the
origin of the long-range terms in the framework of non-equilibrium fluctuating
hydrodynamics.

\section{ Diffusive systems}

        Our purpose in this section is to study the fluctuation dynamics
of purely diffusive systems, which will be characterized by the
number density field, $n(\vec{r}, t)$. According to non-equilibrium
thermodynamics \cite{Lan}, this quantity is governed by the continuity equation

\begin{equation}
\frac{\partial n}{\partial t}+\vec{\nabla} \cdot \vec{J}=0       \label{1i}
\end{equation}

\noindent where $\vec{J}$ is the number density flux, given
by Fick's law

\begin{equation}
\vec{J} =\vec{\vec{D}}\cdot \vec{\nabla}n                      \label{2i}
\end{equation}

\noindent with $\vec{\vec{D}}$ being the diffusion tensor, which becomes a
scalar for isotropic systems. To analyze the fluctuation dynamics, we
will next study two different models with
particular expressions for the diffusion coefficient.

\subsection{Model A}

        As a first model, we will consider diffusion in a system in which the
diffusion coefficient depends on the local number density, as happens,
for example, when hydrodynamic interactions among suspended particles are
considered \cite{Lek}.

        The differential equation governing the evolution of the number density
of particles results from the continuity equation and the linear law for the
mass flux, introduced previously. Since the diffusion coefficient is not
constant, this equation reads

\begin{equation}
\frac{\partial n}{\partial t} = \vec{\nabla}  \cdot \left( D(n)
\vec{\nabla} n \right)
                                                                   \label{2b}
\end{equation}

\noindent where we have taken into account the fact that the diffusion
coefficient depends on position through the local value of the density $n$.
The expressions of both the stationary number density profile
and diffusion coefficient are obtained from the solution of eq.
(\ref{2b}) in the stationary regime

\begin{equation}
\vec{\nabla}\left(D_{s}(n_{s})\cdot\vec{\nabla}n_{s}\right)=0
                                                                \label{2ab}
\end{equation}

\noindent We will analyze the stationary solutions of (\ref{2ab}) in
the situation in which
our system fills the region between two parallel plates and a concentration
gradient $\nabla_{0} n=(n_{L/2}-n_{-L/2})/L \equiv \Delta n/L$ is kept. In this
last expression, $n_{\pm L/2}$ are the values of the number density at the
plates. If the dependence of $D_{s}$ on $n$ is smooth enough, we can expand
both the
stationary values of the diffusion coefficient and the number density in powers
of the parameter $\frac{n_{0}}{D_{0}}\left(\frac{\partial D}{\partial
n}\right)_{s}\equiv \epsilon$, with $n_{0}$ being a characteristic value of the
concentration. Up to linear order in $\epsilon$ one obtains

\begin{equation}
n_{s}(\vec{r}) = n'+\vec{r} \cdot \vec{\nabla}_{0}n +
\frac{1}{2}\frac{\epsilon}{n_{0}} (\vec{r} \cdot \vec{\nabla}_{0}n)^{2}
                                                                \label{4b}
\end{equation}

\noindent  where $n'=\bar{n}-\frac{1}{8} \frac{\epsilon}{n_{0}}
(\Delta n)^{2}$, with $\bar{n}=(n_{L/2}+ n_{-L/2})/2$ being an averaged density
value, and the expression for the diffusion coefficient in this
approximation reads

\begin{equation}
D_{s}(\vec{r})= D_{0}\left[ 1- \epsilon \frac{\bar{n}} {n_{0}} -\frac{\epsilon}
{n_{0}} \vec{\nabla} _{0} n \cdot \vec{r}\right]               \label{5b}
\end{equation}

\noindent Note that for small $\epsilon$, $D_{s}$ remains always
positive. Once we have obtained the stationary solutions, we proceed to
study the
dynamics of the
fluctuations of the number density, $\delta n \equiv n-n_{s}$. According to
fluctuating hydrodynamics, fluctuations follow an equation similar to
(\ref{2b})
with stochastic sources $\vec{J}^{R}(\vec{r},t)$. Note that care should be
taken because the fluctuation dynamics will be affected due to the fact that
 the diffusion coefficient is a fluctuating quantity. After
linearization in the expansion parameter ( an analysis of the case in which
mode-coupling terms are present has been given in \cite{Dia} for a fluid in
stationary flow), and introducing the expression (\ref{5b}), one obtains

\begin{equation}
\frac{\partial \delta n}{\partial t}= D_{0}(1-\epsilon \frac{\bar{n}}{n_{0}})
 \nabla^{2} \delta n - D_{0}\nabla^{2}\left(\frac{\epsilon}{n_{0}} \vec{r}
\cdot  \vec{\nabla}_{0}n\  \delta n \right) - \vec{\nabla} \cdot \vec{J}^{R}
                                                                \label{6b}
\end{equation}

\noindent The random part of the stochastic current satisfies the
fluctuation-dissipation theorem

\begin{equation}
< \vec{J}^{R}(\vec{r},t)~\vec{J}^{R}(\vec{r}^{'},t^{'})> = 2 k_{B} T
\frac{\partial n}{\partial \mu}|_{s}
D_{s}(\vec{r}) \delta (\vec{r}-\vec{r}^{'}) \delta (t-t^{'}) \vec{\vec{1}}
                                                                    \label{7b}
\end{equation}

\noindent where $\vec{\vec{1}}$ is the unit matrix and the average is taken
over a stationary ensemble. Therefore,
$\delta n$ describes a gaussian process \cite{Kei}. In the above expression,
the diffusion coefficient is precisely (\ref{5b}) and the
derivative $\frac{\partial n}{\partial \mu}|_{s}$ must be evaluated in the
steady state. For ideal systems (\ref{7b}) reduces to

\begin{equation}
< \vec{J}^{R}(\vec{r},t)~\vec{J}^{R}(\vec{r}^{'},t^{'})> = 2 k_{B} T
n_{s}(\vec{r}) D_{s}(\vec{r}) \delta (\vec{r}-\vec{r}^{'}) \delta (t-t^{'})
\vec{\vec{1}}                                                   \label{5}
\end{equation}

\noindent which constitutes a good approximation in the case of dilute systems.

         The formal solution for the number density fluctuations,
eq.(\ref{6b}), in Fourier space is given by

\begin{equation}
\delta n(\vec{k},\omega) = -iG(\vec{k}, \omega) \vec{k} \cdot
\vec{J}^{R}(\vec{k}, \omega)
                                                                      \label{7}
\end{equation}

\noindent where we have defined the Green propagator

\begin{equation}
G(\vec{k}, \omega )=\frac{1}{-i\omega + D_{0}\left(1-\epsilon
\frac{\bar{n}}{n_{0}}\right)k^{2} - i \frac{\epsilon}{n_{0}}
k^{2} \vec{\nabla}_{0} n
\cdot \vec{\nabla}_{\vec{k}} }
                                                                   \label{8b}
\end{equation}

\noindent with $\vec{\nabla}_{k}$ being the gradient with respect to the
components of the vector $\vec{k}$.

\noindent Note that in the absence of inhomogeneities ($\epsilon = 0$),
this propagator reduces to

\begin{equation}
G(\vec{k}, \omega)= G_{0}(\vec{k},\omega)=\frac{1}{-i\omega+D_{0}k^{2}}
                                                                \label{9a}
\end{equation}

\noindent which describes the fluctuation dynamics of diffusion processes when
the diffusion coefficient is a constant. The propagator $G(\vec{k},\omega)$ can
also be expanded in powers of $\epsilon$. To linear order one gets

\begin{equation}
G(\vec{k}, \omega)= G_{0}(\vec{k}, \omega) \left[ 1+ \epsilon \frac{D_{0}
\bar{n}}{n_{0}} k^{2} G_{0}(\vec{k},\omega) + i\frac{\epsilon}{n_{0}}
G_{0}(\vec{k},\omega) k^{2} \vec{\nabla}_{0}n \cdot
\vec{\nabla}_{\vec{k}} -2 i \epsilon \frac{D_{0}}{n_{0}}
G_{0}(\vec{k},\omega)^{2}
\vec{\nabla}_{0}n \cdot \vec{k}\right]                            \label{9}
\end{equation}

\noindent where we have considered that $G(\vec{k},\omega)$ is an operator.
This expression will be used in the next section to compute correlation
functions.

\subsection{Model B}

        We now consider a different model, in which the diffusion
process takes place in an
inhomogeneous medium in which the diffusion
coefficient may depend on the position according to the expression

\begin{equation}
D(\vec{r})= D_{0}+\vec{\nabla} _{0}D \cdot \vec{r}    \label{1}
\end{equation}

\noindent Here $D(\vec{r})$ is the diffusion coefficient, $D_{0}$ is a
constant and  $\vec{\nabla}_{0} D$ is a constant vector that will be
determined by boundary conditions. Variations of the diffusion coefficient are
limited to ensure the positive character of such a quantity. The former
expression could be interpreted as an expansion in the parameter
$L\frac{|\vec{\nabla}_{0} D|}{D_{0}}$, with $L$ being a characteristic
length of the
system, in which higher order powers of the parameter have been neglected.

        The one-dimensional microscopic version of this model could be a
particle performing a random walk in the presence of traps whose distribution
depends on position \cite{Kam}. It is possible to show that the probability
distribution satisfies a Fokker-Planck equation in which one may identify the
position-dependent diffusion coefficient as a function of the probability
distribution of the traps.  If this distribution varies slowly with the
position, one
obtains an expression similar to (\ref{1}).

        As in model A, the number density evolves according to

\begin{equation}
\frac{\partial n}{\partial t} = \vec{\nabla}  \cdot D(\vec{r}) \vec{\nabla}  n
                                                                    \label{2}
\end{equation}

\noindent We will analyze the stationary solutions of (\ref{2}) subjected to
the
same boundary conditions as in the previous model. To linear order in
$\Delta D/D_{0}$, we get

\begin{equation}
n_{s}(\vec{r}) = \tilde{n}+\vec{r} \cdot \vec{\nabla}_{0}n - \frac{1}{2
D_{0}}\vec{r}
 \cdot \vec{\nabla}_{0}n\left( \vec{r} \cdot \vec{\nabla}_{0} D\right)
                                                                     \label{3}
\end{equation}

\noindent where we have defined $\tilde{n}\equiv \bar{n}-\frac{1}{8}
\frac{\Delta n \Delta D}{D_{0}}$.

        We now proceed to study the dynamics of the number density
fluctuations. Note that fluctuation dynamics in this model will differ from
(\ref{6b}) because now the diffusion coefficient does not fluctuate. To linear
order in the expansion parameter one obtains

\begin{equation}
\frac{\partial \delta n}{\partial t}= D_{0}\nabla^{2} \delta n+\vec{r} \cdot
\vec{\nabla}_{0}D \nabla^{2} \delta n + \vec{\nabla}_{0}D \cdot \vec{\nabla}
\delta n - \vec{\nabla} \cdot \vec{J}^{R}                           \label{6}
\end{equation}

\noindent where use has been made of eq.(\ref{1}). Here, $\vec{J}^{R}$ is again
the
stochastic flux which satisfies the fluctuation-dissipation theorem formulated
in eq.(\ref{5})  since the system is ideal. In it, the diffusion coefficient is
now given by eq.(\ref{1}).

        The formal solution for the number density fluctuations in Fourier
space  also follows from eq.(\ref{7}), with the appropriate Green function,
which now reads

\begin{equation}
G(\vec{k}, \omega )=\frac{1}{-i\omega + D_{0}k^{2} + i k^{2} \vec{\nabla}_{0}D
\cdot \vec{\nabla}_{\vec{k}} + i \vec{\nabla}_{0}D \cdot \vec{k}}
                                                                   \label{8}
\end{equation}

        This expression differs from (\ref{8b}) because now the diffusion
coefficient does
not fluctuate. In the same way as we have done in model A, the propagator can
be expanded in powers of  $\frac{\Delta D}{D_{0}}$.

\section{ Correlation Functions}

        To compute the number density correlation function, for model A, we
will use
(\ref{7}) and the expression for the fluctuation-dissipation theorem (\ref{5}),
in ($\vec{k}$,$\omega$) representation. We then obtain

\begin{eqnarray}
<\delta n(\vec{k},\omega) \delta n(\vec{k}^{'},\omega^{'})>&=&-
G(\vec{k},\omega) G(\vec{k}^{'},\omega^{'}) \vec{k} \cdot \vec{k}^{'}\left[
D_{0} n' + i n'\epsilon \nabla_{0}n \cdot \vec{\nabla}_{k} \right. \nonumber
\\&+& \left. i
D_{0}\epsilon \nabla_{0}n \cdot \vec{\nabla}_{k} -\frac{1}{2}
\vec{\nabla }_{0}n \vec{\nabla}_{0}n:\vec{\nabla }_{k} \vec{\nabla
}_{k}\right] \delta(\vec{k}+\vec{k}^{'})
                                                                 \label{10}
\end{eqnarray}

\noindent where, due to the convolution in real space, $G(\vec{k}, \omega)$
should be understood as an operator. The equal-time correlation function in
real space follows after Fourier transforming (\ref{10}) according to

\begin{eqnarray}
<\delta n (\vec{r},t), \delta n(\vec{r}^{'},t)>&=&
\frac{1}{(2\pi)^{8}}
\int_{-\infty}^{\infty} d\vec{k} e^{i \vec{k}\cdot\vec{r}}
\int_{-\infty}^{\infty} d\vec{k}^{'} e^{i\vec{k}^{'}\cdot\vec{r}^{'}} \cdot
						\nonumber \\
&&\left\{ \int_{-\infty}^{\infty} d\omega e^{-i\omega t}
\int_{-\infty}^{\infty} d\omega^{'} e^{-i \omega^{'} t}
<\delta n(\vec{k},\omega) \delta n(\vec{k}^{'},\omega^{'})> \right\}
                                                           \label{11}
\end{eqnarray}

\noindent As we have a finite system, one should have used Fourier series in
the spatial direction in which the external gradient is applied. This fact
means that our expressions will be restricted for points $\vec{r}$ and
$\vec{r}^{'}$ such that $|\vec{r}-\vec{r}^{'}|<<2 L/\pi$, with $L$ being the
size of the system. Otherwise, the discrete character of the reciprocal space
has to be taken into account \cite{Tor}. Performing the integrals, we finally
arrive at

\begin{equation}
<\delta n(\vec{r},t) \delta n(\vec{r}^{'},t)>= n_{s}(\vec{r})
\delta(\vec{r}-\vec{r}^{'}) + \frac{\Gamma(d-2)}{2^{d}
\pi^{\frac{d-2}{2}}\Gamma(\frac{d}{2})}
\frac{\epsilon}{n_{0}}\frac{|\vec{\nabla}_{0}n|^{2}}
{{|\vec{r}-\vec{r}^{'}|^{d-2}}}
                                                          \label{corr1}
\end{equation}

\noindent to linear order in $\epsilon$. In the former expression $\Gamma
(x)$ is the Euler function, and it has been given for dimensions d$>$2. When
d=2 logarithmic divergencies appear. As expected, eq.(\ref{corr1}) reduces to
its equilibrium counterpart when the gradient is turned off.

        For the sake of simplicity, the long-range contribution in
eq.(\ref{corr1}) has been obtained keeping only the first term in the virial
expansion of the derivative $\partial n/\partial\mu$, which corresponds to the
ideal case. Potential interactions between particles
would introduce additional terms in $\partial n/\partial\mu$ \cite{Lek}, but
long-range behavior is still present.

        Following the same line of reasoning, we can also compute the
correlation function for model B introduced in the previous section. Up linear
order in $|\nabla_{0}D|$, we arrive at the stationary function

\begin{equation}
<\delta n(\vec{r},t) \delta n(\vec{r}^{'},t)>= n_{s}(\vec{r})
\delta(\vec{r}-\vec{r}^{'})                     \label{corr2}
\end{equation}

\noindent which does not exhibit long-range correlations. This expression
differs from the equilibrium result since now the local number density enters
into the correlation. Equation (\ref{corr1}) reduces to this expression as
well, if a constant diffusion coefficient is considered.

        At this point it is worth comparing the above results with the exact
ones obtained for a thermal diffusive system \cite{Agu}. Although both
processes are diffusive and consequently, at the deterministic level, they are
described by the same equations, one should emphasize the different behavior
between concentration and temperature correlation functions in the presence of
external gradients. Indeed, while the former does not always exhibit long-range
behavior, in the latter long-range correlations are always present. The
equal-time temperature correlation function has been found to be

\begin{equation}
<\delta T(\vec{r},t) \delta T(\vec{r}^{'},t)>= \frac{k_{B}}{\rho c_{v}}
\left\{ T_{s}^{2}(\vec{r}) \delta(\vec{r}-\vec{r}^{'}) +
\frac{|\vec{\nabla}_{0}T|^{2}}{4 \pi |\vec{r}-\vec{r}^{'}|} \right\}
                                                                    \label{24}
\end{equation}

\noindent where $T_{s}(\vec{r})$ is the stationary linear temperature profile,
$\vec{\nabla}_{0}T$ the external temperature gradient, $k_{B}$ the Boltzmann's
constant, $\rho$ the density and $c_{v}$ the specific heat at
constant volume. Note that eqs.(\ref{corr1}) and (\ref{24}) have the
same form, which indicates that the modifications in the propagator
induced by the inhomogeneities in the diffusion coefficient do not
modify the long-range behavior of the correlations.

	From expression (\ref{corr1}) it is straightforward to compute the static
structure factor, which for d=3 is

\begin{equation}
S(\vec{k})=n_{s}(\vec{k})+\frac{\epsilon}{n_{0}}
\frac{|\nabla_{0}n|^{2}}{k^2}
                                         \label{sf}
\end{equation}

\noindent An estimation of the non-local contribution, $S_{nl}$, of
eq.(\ref{sf}), given through the second term of its right hand side, relative
to its equilibrium value, $S_{eq}$, follows from

\begin{equation}
\frac{S_{nl}(k)}{S_{eq}(k)}= \frac{\epsilon |\nabla_{0}n|^{2}}{n_{0}^{2} k^{2}}
                                               \label{sfc}
\end{equation}

\noindent  In fluid systems, the concentration gradient will induce a coupling
between concentration and velocity fluctuations \cite{Seg}. This fact leads to
a non-local term, proportional to $k^{-4}$, which is more important than the
above mentioned correction. However, if the wave vector is parallel to the
external concentration gradient this coupling may be avoided \cite{Pag}. For
this geometry, $S_{nl}$ given through eq.(\ref{sfc}), is the only non-local
correction to the static structure factor. For a colloidal suspension of hard
spheres the quantity (\ref{sfc}) is proportional to the volume fraction
\cite{Lek} and to the ratios $\Delta n/n_{0}$ and $(k L)^{-1}$.

\section{Discussion}

         We have shown that purely diffusive systems
away from equilibrium may exhibit long-range correlations. In order to
clarify the origin of such correlations we will first analyze in detail the
nature of the Langevin sources.

        Our starting point is the expression for the entropy
production corresponding to a irreversible vectorial process, which
is given as a product of a flux-force pair \cite{Gro}. Denoting the flux by
$\vec{J}$
and the force by $\vec{X}$, one has

\begin{equation}
\sigma =\vec{J} \cdot \vec{X}                                      \label{15}
\end{equation}

\noindent from which one formulates the linear law

\begin{equation}
\vec{J}=\vec{\vec{L}} \cdot \vec{X}                                \label{16}
\end{equation}

\noindent where $\vec{\vec{L}}$ is the matrix of phenomenological coefficients.

        In the framework of fluctuating hydrodynamics, fluctuations are
incorporated simply by adding stochastic sources to the currents. One has

\begin{equation}
\vec{J} =\vec{\vec{L}} \cdot \vec{X} + \vec{J}^{R}                 \label{17}
\end{equation}

\noindent where the stochastic current, denoted in general by $\vec{J}^{R}$,
introduces a gaussian
white noise stochastic process with zero mean and whose correlations are given
by

\begin{equation}
<\vec{J}^{R}(\vec{r},t)~\vec{J}^{R}(\vec{r}^{'},t^{'})>= \vec{\vec{\Gamma}}
\delta(\vec{r}-\vec{r}^{'}) \delta (t-t^{'})                       \label{18}
\end{equation}

\noindent The matrix $\vec{\vec{\Gamma}}$ is related to the matrix of
phenomenological coefficients by means of the relation
$\vec{\vec{\Gamma}}=2k_{B}\vec{\vec{L}}$. This last expression, together with
eq.(\ref{18}), constitutes the formulation of the fluctuation-dissipation
theorem.

        Note that, according to fluctuating hydrodynamics, the former scheme
holds when
fluctuations occur around equilibrium or nonequilibrium steady states. In this
last case,
the matrix  $\vec{\vec{\Gamma}}$ may depend on the position since the matrix of
phenomenological coefficients may depend on local equilibrium quantities to
ensure local gaussianity. As a first example, let us consider the diffusion of
a contaminant. In this case the entropy production reads

\begin{equation}
\sigma =\vec{J}_{D} \cdot \vec{\nabla} \frac{\mu}{T}                 \label{19}
\end{equation}

\noindent from which one derives the linear law

\begin{equation}
\vec{J}_{D}=\frac{1}{T}\vec{\vec{L}}_{D} \cdot \vec{\nabla } \mu  \label{20}
\end{equation}

\noindent In these two last expressions, $\vec{J}_{D}$ is the diffusion
current, $\mu$ the chemical potential, $T$ the temperature (assumed to be
constant) and $\vec{\vec{L}}_{D}$ the corresponding matrix of phenomenological
coefficients.  The linear law (\ref{20}) can be identified with Fick's law,
$\vec{J}_{D}=-\vec{\vec{D}} \cdot \vec{\nabla }n$, where $\vec{\vec{D}}
=\frac{1}{T}\vec{\vec{L}}_{D}\frac{\partial \mu}{\partial n}$.  This
identification enables us to formulate the fluctuation-dissipation
theorem \cite{Kei}

\begin{equation}
\vec{\vec{\Gamma}}=2k_{B} T \vec{\vec{D}} \frac{\partial n}{\partial \mu}
                                                                   \label{21}
\end{equation}

\noindent which for non-interacting particles becomes

\begin{equation}
\vec{\vec{\Gamma}} = 2\vec{\vec{D}}n                          \label{22}
\end{equation}

        One can also apply our former analysis to the case of heat conduction.
One arrives at

\begin{equation}
\vec{\vec{\Gamma}} = 2k_{B}T^{2} \vec{\vec{\lambda}}                \label{23}
\end{equation}

\noindent where $\vec{\vec{\lambda}}$ is the thermal conductivity related to
the corresponding matrix of phenomenological coefficients $\vec{\vec{L}}_{q}$
through $\vec{\vec{\lambda}}=\frac{1}{T^{2}}\vec{\vec{L}}_{q}$.

         In the case of heat conduction, for a constant thermal
conductivity, the quadratic dependence of the
matrix $\vec{\vec{\Gamma}}$ on temperature is responsible for the long-range
behavior of the static correlation function shown in (\ref{24}). Concerning
mass diffusion, long-range correlations are also present only when the
corresponding matrix $\vec{\vec{\Gamma}}$ is quadratic in the thermodynamic
field. In fact, in model A one has a quadratic dependence of
$\vec{\vec{\Gamma}}$ on $n$ owing to the dependence of the diffusion
coefficient
on the local density field. In model B the diffusion coefficient depends on
position, but not on the density, and therefore the matrix $\vec{\vec{\Gamma}}$
is linear in $n$.

        In order to give a simple justification of the long-range behavior, we
may consider the following intuitive argument proposed by Ronis\cite{Trem}. If
we have a
single Langevin
source at a certain point generating randomly hydrodynamic excitations, it will
strongly correlate, at equal times, two equidistant points which are closer to
this source than the typical hydrodynamic decay length. In equilibrium, as we
have uncorrelated Langevin sources everywhere, all of them being equivalent in
intensity and magnitude, the correlations between these two points will drop
out. Out of equilibrium, however, as the Langevin forces are
detuned, long-range correlations could in principle appear.

        In the models we have analyzed, while the Langevin forces are detuned
long-range correlations are not always present. Consequently, the previous
argument should be somehow precised. In model A, and also in thermal diffusive
systems, the ratio between the intensity of the noise $\vec{\vec{\Gamma}}$ and
the local value of the corresponding field is proportional to the stationary
field, whereas in model B it does not depend on it. It is our contention that
long-range behavior originates from the nature of the Langevin sources, which
should increase their intensity relative to the local thermodynamic
field, and not only from their detuned character.  Moreover, the
form of eq.(\ref{corr1}) indicates that the modifications in the
propagator due to the inhomogeneities in the diffusion coefficient
and its fluctuations do not contribute to the long-range behavior of
the correlations, which appear only through the above-named mechanism.
Finally, note that the correlation functions decay as $k^{-2}$ as we have
purely diffusive systems and there is no coupling with the fluctuations of
 other hydrodynamic fields.

        Long-range correlations have also been predicted in other
nonequilibrium
systems, as for example for fluids under temperature gradients\cite{Ron}, for
diffusive systems in the presence of chemical reactions\cite{Kei} or in
semiconductors in the presence of an electric field\cite{Gun}. In fluid
systems, correlations between density or temperature and velocity appear out of
equilibrium owing to the breaking of time reversal symmetry and have been
observed
experimentally in Brillouin scattering \cite{Bey}. What regards
Rayleigh scattering, both in simple fluids \cite{Law} and in binary mixtures
\cite{Seg}, long-range correlations appear due to the resonant coupling between
temperature or concentration fluctuations and transverse velocity fluctuations
\cite{Ro2}, which leads to the characteristic $k^{-4}$ behavior of the static
structure factor. In the remaining examples, long-range behavior is related to
the existence of characteristic frequencies or lengths. Note that in our case,
as in thermal diffusion, detailed balance holds even in the presence of
external constraints\cite{Nic}.

         Contrasting with our analysis, other diffusive systems have been
considered in the literature, which do not satisfy a fluctuation-dissipation
theorem,  either because they come from an external noise source
\cite{Grn} or because the internal noise does not have a thermal origin
\cite{For}-\cite{Gar}. In all these examples, the appearance of long-range
correlations typically originate from some kind of essential anisotropy.
However, when fluctuation-dissipation is satisfied, anisotropy does not
guarantees long-range order.  In fact, it is easy to see that model A with a
constant but anisotropic diffusion matrix does not produce long-range
correlations.

        Model A is a real hydrodynamic
model in which long-range correlations originate from hydrodynamic interactions
between particles. A similar model has been proposed as the hydrodynamic limit
of a lattice gas of interacting particles \cite{Spo}.  Though a constant
diffusion coefficient is considered, the interaction introduces nonlinearities
in the number density in the fluctuation-dissipation theorem due to the
nonideal character of the system.  The long-range term that is obtained can be
deduced from our model A if $\epsilon$ is set equal to zero and the next order
in the virial expansion of $\partial n / \partial \mu $ is taken into account.

\acknowledgements

         This work has been supported by CICyT, grant PB92-0859. One of
us (I. P.) wants to thank Ministerio de Educaci\'{o}n y Ciencia for
financial support.

\end{document}